# A time-space Hausdorff fractal model for non-Fickian transport in porous media

Yingjie Liang[1], Ninghu Su[2,3,4], Wen Chen[1], Xu Yang[1]

[1]Institute of Soft Matter Mechanics, College of Mechanics and Materials, Hohai University, Nanjing, Jiangsu 211100, China

[2]College of Science and Engineering and [3]Tropical Water and Aquatic Ecosystem Research, James Cook University, Cairns, Queensland 4870, Australia

[4]College of Resources and Environmental Sciences, Ningxia University, Yinchuan, Ningxia 750021, China

**Corresponding author**: Yingjie Liang (liangyj@hhu.edu.cn)

**Abstrac**t: This paper presents a time-space Hausdorff derivative model for depicting solute transport in aquifers or water flow in heterogeneous porous media. In this model, the time and space Hausdorff derivatives are defined on non-Euclidean fractal metrics with power law scaling transform which, respectively, connect the temporal and spatial complexity during transport. As an example of applications of this model, an explicit solution with a constant diffusion coefficient and flow velocity subject to an instantaneous source is derived and fitted to the breakthrough curves of tritium as a tracer in porous media. These results are compared with those of a scale-dependent dispersion model and a time-scale dependent dispersion model. Overall, it is found that the fractal model based on the Hausdorff derivatives better captures the early arrival and heavy tail in the scaled breakthrough curves for variable transport distances. The estimated parameters in the fractal Hausrdorff model represent clear physical mechanisms.

**Keyword**s: Time-space dependent dispersion, Hausdorff derivative, fractal PDE, solute transport, water flow, porous media

## 1. Introduction

The terminologies of solute and porous media cover two very large categories of natural and industrial materials. Solute is a generic term for particles, granules or microbes in nature such as salts in geological strata, nutrients for plants, and also agro-chemicals in agriculture *etc*. Porous media is a another generic term used to cover a very wide range of materials such as polymers and powders in industries, tissues in biological bodies, and soils and rocks commonly on the Earth and other astronomical bodies. Solute and porous media are two topics with great importance for geo-environmental, industrial and agricultural practices, and their quantification attracts growing interest, particularly on issues such as solute transport in heterogeneous media (Chang and Yeh, 2016; Willson et al., 2006; Moslehi et al., 2016; Lee et al., 2017) which is related to pollutants in soils and aquifers, and nutrients for plants. A major feature of solute transport in heterogeneous media is the scale-dependent dispersion (Gao et al., 2010; Deng and Jung, 2009; Mishra and Parker, 2010; Pedretti et al., 2016; Sabahi et al., 2015), *i.e.*, the flow velocity and the dispersion coefficient are not constant values, and instead they vary as a function of time, distance or both. Such scale-dependent transport behaviors are also referred to as anomalous (Dentz et al., 2004), non-Gaussian (Uffink et al., 2012) or non-Fickian transport (Bijeljic et al., 2011).

It is well known that the classical advection-dispersion model (Thongmoon et al., 2012) with a constant dispersion coefficient and flow velocity fails to model the non-Fickian transport. To remedy this problem, many sophisticated models (*e.g.*, the time- and space-dependent dispersion model (Su et al., 2005), the continuous-time random walk (CTRW) model (Dentz et al., 2015), the fractional derivative dispersion model (Sun et al., 2014), and the Hausdorff derivative dispersion model (Cai et al., 2017) have been developed to quantify the observed non-Fickian transport. However, the vast majority of the time- and space-dependent dispersion models published to date are empirical or semi-empirical (Barry and Sposito, 2005) in which the dispersion coefficient or the flow velocity is either a function of time or distance except for very few of them as a function of both time and space (Su et al., 2005). The fractional derivative model in the form of fractional partial differential equations (FPDEs) can be derived by several approaches including the CTRW theory, and its fractional orders of differentiation quantitatively connect the complexity of temporal and spatial changes in heterogeneous media (Metzler and Klafter, 2000; Su, 2014). The main drawbacks of the fractional derivative model are its computational expense and it cannot directly capture the geometric information of the heterogeneous media (Chen et al., 2010) since the FPDEs published so far were derived in the Euclidean space. Compared with the fractional derivative, the Hausdorff derivative (Chen, 2006) is a simpler and computationally efficient tool with a clear mathematical and physical definition to incorporate the structural complexity and geometrical information into the orders of time and space derivatives. In this paper, a new Hausdorff fractal model is presented to characterize the non-Fickian solute transport in heterogeneous porous media.

The Hausdorff derivative, *i.e.*, the Hausdorff fractal derivative, was first proposed by Chen (2006), and is defined on the non-Euclidean fractal metrics using a time-space scaling transform (Chen et al., 2017). From the perspective of statistics, a fundamental solution of the space and time Hausdorff derivative transport equations appears as a stretched Gaussian distribution or a stretched exponential function (Chen 2006). The definition of the Hausdorff derivative on a fractal geometry is consistent with the one given by Li and Ostoja-Starzewski (2009) where the space dimensionality is the dimension of the fractal. In recent years, the Hausdorff derivative has attracted great attention in diverse fields such as magnetic resonance imaging (Lin, 2017), creep of viscoelastic materials (Cai et al., 2016), non-Newtonian fluid dynamics (Su et al., 2017), and heat conduction (Reyes-Marambio et al., 2016), but only a small number of reports in the field of solute transport such as in (Sun et al., 2017).

Recently Allwright and Atanaga (2018) developed a space Hausdorff derivative advection dispersion equation, which was used for simulating the effects of dynamic fractal dimension in a one-dimensional groundwater flow system. Nie *et al*. (2017) reported an explicit solution of a general Hausdorff derivative model for the transport of the suspended sediment in unsteady flows. Liu *et al*. (2017) established a variable order fractal derivative model. Sun *et al*. (2013) proposed a fractal Richards' equation to capture the water movement in unsaturated media with non-Boltzmann scaling. Liang *et al*. (2016) used the fractal derivative model in the spatial frequency domain to characterize the anomalous diffusion of water in heterogeneous media. However, the physical interpretations of the parameters in the Hausdorff derivative models are not very clear, and their feasibility in depicting solute transport in porous media is yet to be tested.

In this study, a time-space Hausdorff fractal advection and dispersion model is presented by introducing the Hausdorff time and space derivatives into the Fick's second law for water movement in heterogeneous porous media, and with published data on the tritium movement in porous media as an example. An explicit solution of this model is presented for an instantaneous source and compared with the similarity solution of a non-fractal time-scale-dependent dispersion model (Su et al., 2005). The relationships between the parameters in the Hausdorff fractal model and the transport distance are also investigated to provide physical interpretations of the proposed model.

This paper is organized as follows: in Section 2, the time and space Hausdorff fractal advection and dispersion model is presented and its solution subject to an instantaneous source solution is derived. Section 3 describes the experiment data and the methods used to fit the data to derive the model parameters. Section 4 provides the results in terms of the scaled breakthrough curves, comparable fitting and interpretations of the parameters for different cases. The results are discussed in Section 5, and finally, in Section 6 some conclusions are drawn.

## 2 Theory

### 2.1. The advection-dispersion equation of solute transport in porous media

We start with the equation of solute transport in porous media, which, without considering various sources, is of the form given by Su *et al.* (2005)

$$\frac{\partial c(x,t)}{\partial t} = \frac{\partial}{\partial x}\left[D(x,t)\frac{\partial c(x,t)}{\partial x}\right] - \frac{\partial}{\partial x}[Vc(x,t)] \qquad (1)$$

where $c(x,t)$ is the solute concentration, i.e., the probability density function of the propagator located in an infinitesimal neighborhood $dx$ centered in the distance $x$ at time $t$; $D$ is the diffusion (or dispersion) coefficient, and $V$ is the flow velocity which could also be a function of the travel distance and time.

An extensive mathematical analysis of Eq. (1) with various solutions subject to different types of initial and boundary conditions can be found in (Su et al., 2005) where the data from field experiments were used to verify some of the solutions. In this paper we combine Chen's concept of temporal-spatial Hausdorff fractal media (Chen et al., 2006) into Eq. (1), derive an explicit solution of the new model and verify it with the data from the field (Pang and Hunt, 2001).

## 2.2. The Hausdorff fractal derivatives and the resultant fractal governing equation

The Hausdorff derivative and its corresponding advection-dispersion equation are defined in the context of fractal metrics proposed by Chen (2006). In one dimensional topological fractal media, the metric transform is defined as

$$\begin{cases} \Delta \hat{t} = \Delta t^{\alpha} \\ \Delta \hat{x} = \Delta x^{\beta} \end{cases} \tag{2}$$

where $\Delta x$ and $\Delta t$ denote the distance interval and time intervals, respectively [20], $\alpha$ and $\beta$ represent the fractal dimensionality in time and space, respectively. These metric transforms satisfy two hypotheses: the fractal invariance and the fractal equivalence. The above definition in Eq. (2) is a special case given by Balankin and Elizarraraz (2012).

In the context of fractal metrics, the distance a particle travelled can be expressed as the length along which the particle moved under the fractal time (Chen et al., 2017),

$$s(t) = v(t - t_0)^{\alpha} \tag{3}$$

where $s(t)$ denotes the distance, $v$ the uniform velocity, $t$ the current time instance, $t_0$ the initial instance. When the velocity fluctuates with time, the Hausdorff integral distance is determined as

$$s(t) = \int_{t_0}^{t} v(\rho) d(\rho - t_0)^{\alpha} \tag{4}$$

where $\rho$ is the motion time. Then the velocity on fractals is defined based on Eq. (4)

$$\frac{ds(t)}{dt^{\alpha}} = \lim_{t_1 \to t} \frac{s(t_1) - s(t)}{(t_1 - t_0)^{\alpha} - (t - t_0)^{\alpha}} = \frac{1}{\alpha(t - t_0)^{\alpha-1}} \frac{ds(t)}{dt} \tag{5}$$

and when the initial instance $t_0 = 0$, Eq. (5) degenerates into

$$\frac{ds(t)}{dt^{\alpha}} = \lim_{t_1 \to t} \frac{s(t_1) - s(t)}{t_1^{\alpha} - t^{\alpha}} = \frac{1}{\alpha t^{\alpha-1}} \frac{ds(t)}{dt} \tag{6}$$

thus, the Hausdorff derivatives of the solute concentration $c(x,t)$ in time with the initial instance $t_0 = 0$, and space with the initial position $x_0 = 0$, can be defined, respectively, as

$$\frac{dc(x,t)}{dt^{\alpha}} = \lim_{t_1 \to t} \frac{c(x,t_1) - c(x,t)}{t_1^{\alpha} - t^{\alpha}} = \frac{1}{\alpha t^{\alpha-1}} \frac{dc(x,t)}{dt} \tag{7}$$

and

$$\frac{dc(x,t)}{dx^\beta} = \lim_{x_1 \to x} \frac{c(x_1,t) - c(x,t)}{x_1^\beta - x^\beta} = \frac{1}{\beta x^{\beta-1}} \frac{dc(x,t)}{dx} \qquad (8)$$

It should be pointed out that compared with the fractional derivatives, the above defined Hausdorff derivatives do not involve an integral convolution and are local in nature. Such definitions of derivatives in Eqs. (7) and (8) are termed *fractal derivatives* as opposed to the definitions of *fractional derivatives*.

Based on the Hausdorff derivatives in Eqs. (7) and (8), the time-space advection-dispersion equation (ADE) in Eq. (1) is written [Chen, 2006]

$$\frac{\partial c(x,t)}{\partial t^\alpha} = \frac{\partial}{\partial x^\beta}\left(D(x,t)\frac{\partial c(x,t)}{\partial x^\beta}\right) - \frac{\partial}{\partial x^\beta}[Vc(x,t)] \qquad (9)$$

which can be written in the usual form of the PDE,

$$\frac{\partial c(x,t)}{\partial t} = \frac{\alpha t^{\alpha-1}}{\beta x^{\beta-1}} \frac{\partial}{\partial x}\left(\frac{D(x,t)}{\beta x^{\beta-1}}\frac{\partial c(x,t)}{\partial x}\right) - \frac{\alpha t^{\alpha-1}}{\beta x^{\beta-1}} \frac{\partial}{\partial x}[Vc(x,t)] \qquad (10)$$

The diffusion (or dispersion) coefficient in Eq. (9) was given by Su *et al.* (2005) as

$$D(x,t) = D_0 x^m t^\lambda \qquad (11)$$

where $m$ and $\lambda$ are fractal parameters of the media and flow, respectively. There are options for the further parameterization of Eq. (10) by employing the spatial fractal dispersivity by Wheatcraft and Tyler (1988), and combining Chen's (2006) concept of fractal space as indicated by Eq. (2). The Wheatcraft-Tyler model for the dispersivity, $\alpha_m$, which is equivalent to $D_0 x^m$ with $\lambda = 0$ in Eq. (11), is of the form

$$\alpha_m = D_0 x^m = \frac{1}{2} D_0 \sigma^2 x^{2d-1} \qquad (12)$$

which relates the dispersivity to the straight-line Euclidean distance, $x$, along which the particles travel with variance, $\sigma^2$, and fractal dimension of $d$. In this paper we employ Chen's fractal concept defined in Eq. (2) which is regarded as being equal to $\beta$ resulting in a fractal length $x^\beta$, and similarly for a fractal time $t^\alpha$. Then, by combining Eqs. (1), (2), (11) and (12) with $d = \beta$ and $\lambda = \alpha - 1$, one arrives at

$$D(x,t) = \frac{D_0 \sigma^2}{2} x^{\beta(2\beta-1)} t^{\alpha(\alpha-1)} \tag{13}$$

With $c = c(x,t)$ for simplicity and substitution of Eq. (13) in Eq. (10) yields

$$\frac{\partial c}{\partial t} = \frac{\alpha D_0 t^{\alpha-1}}{2\beta^2 x^{\beta-1}} \frac{\partial}{\partial x}\left(\sigma^2 x^{2\beta^2-1} t^{\alpha^2-\alpha} \frac{\partial c}{\partial x}\right) - \frac{\alpha t^{\alpha-1}}{\beta x^{\beta-1}} \frac{\partial}{\partial x}(Vc) \tag{14}$$

The case of $\lambda = \alpha - 1$ with $\alpha \in (0,1]$, i.e., $0 < \alpha \leq 1$, is consistent with the case for time-dependent dispersion only (Muralidhar and Ramkrishna, 1993). Eq. (14) is the time-space fractal advection-dispersion equation, which is a further extension to a spatial fractal ADE with a constant dispersion coefficient by Allwright and Atangana (2018).

Fractal media in terms of spatial variability is easy for practitioners such as hydrologists and soil scientists to understand while fractal media in terms of fractal time is less comprehensible. For this reason, here we only investigate a simplified form of Eq. (14) as an example. With a constant velocity and constant $\sigma^2$ with $\alpha = 1$ Eq. (14) simplifies as

$$\frac{\partial c}{\partial t} = \frac{D_0 \sigma^2}{2\beta^2 x^{\beta-1}} \frac{\partial}{\partial x}\left(x^{2\beta^2-1} \frac{\partial c}{\partial x}\right) - \frac{V}{\beta x^{\beta-1}} \frac{\partial c}{\partial x} \tag{15}$$

which is a result of the fact that, for a given media, the variance $\sigma^2$ is constant.

In a compact form Eq. (15) is written as

$$\frac{\partial c}{\partial t} = A x^{1-\beta} \frac{\partial}{\partial x}\left(x^\eta \frac{\partial c}{\partial x}\right) - \frac{V x^{1-\beta}}{\beta} \frac{\partial c}{\partial x} \tag{16}$$

with

$$A = \frac{D_0 \sigma^2}{2\beta^2} \tag{17}$$

and

$$\eta = 2\beta^2 - 1 \tag{18}$$

Eq. (16) can be expanded as

$$\frac{\partial c}{\partial t} = Ax^{2\beta^2-\beta}\frac{\partial^2 c}{\partial x^2} + \left(A\eta x^{\eta-\beta} - \frac{Vx^{1-\beta}}{\beta}\right)\frac{\partial c}{\partial x} \tag{19}$$

Eq. (19) is similar to the generalized Feller equation (GFE) investigated by Lehnigk (1993) except for the term $\frac{Vx^{1-\beta}}{\beta}$ where the GFE has a linear term in the form of $\varepsilon x$ with $\varepsilon$ constant.

We have presented a spatial Hausdorff fractal PDE for solute transport in aquifers in Eq. (16) which is equivalent to Eq. (19). In the following section we will present one form of the solutions with a constant diffusion coefficient and constant velocity, and verify its application with the data.

## 3. Solutions of the Hausdorff fractal PDE with a constant diffusion coefficient and constant flow velocity

With $c = c(x,t)$ for simplicity and the pair of transformations

$$\begin{cases} \hat{t} = t^\alpha \\ \hat{x} = x^\beta \end{cases} \tag{20}$$

Eq. (9) becomes

$$\frac{\partial c}{\partial \hat{t}} = D\frac{\partial^2 c}{\partial \hat{x}^2} - V\frac{\partial c}{\partial \hat{x}} \tag{21}$$

which, with the aid of the Fuerth transform (Jost, 1960),

$$c = u\exp\left(\frac{V\hat{x}}{2D} - \frac{V^2\hat{t}}{4D}\right) \tag{22}$$

can be transformed to the standard diffusion equation or heat equation,

$$\frac{\partial u}{\partial \hat{t}} = D\frac{\partial^2 u}{\partial \hat{x}^2} \tag{23}$$

With an instantaneous source input, *i.e.*, a Dirac delta function input, $\delta(\hat{x})$,

$$u = M\delta(\hat{x}), \quad \hat{t} = 0 \tag{24}$$

$$u \to 0, \quad \hat{t} \to \infty \tag{25}$$

where the total material released at the origin is

$$M = \int_0^\infty c d\hat{x} \qquad (26)$$

the solution of Eq. (23) is well-known such as in (Crank, 1975)

$$u(\hat{x},\hat{t}) = \frac{M}{(\pi D \hat{t})^{1/2}} \exp\left(-\frac{\hat{x}^2}{4D\hat{t}}\right) \qquad (27)$$

which, by reinstating the original variables in Eqs. (20) and (22), becomes

$$c(x,t) = \frac{M}{(\pi D t^\alpha)^{1/2}} \exp\left[\frac{1}{2D}\left(Vx^\beta - \frac{x^{2\beta}}{2t^\alpha} - \frac{V^2 t^\alpha}{2}\right)\right] \qquad (28)$$

Eq. (28) is the solution of the Hausdorff time-space fractal model with a constant diffusion coefficient and constant velocity. The dimensions of $D$ and $V$ in Eq. (28) are, respectively, $[L^{2\beta}/T^\alpha]$, and $[L^\beta/T^\alpha]$, where $L$ is the unit of length and $T$ is the unit of time. Bear in mind that Eq. (28) is the solute concentration in the Euclidean space with the Hausdorff fractal space-time parameters, $\beta$ and $\alpha$.

The connections between the Hausdorff fractal space and time are given by Eq. (2), and to express the dimensions of $D$ and $V$ in Eq. (28) in the Hausdorff fractal space and time, Eq. (2) is used to yield the dimensions of $D$ and $V$ as $[D]=[L^2/T]$ and $[V]=[L/T]$. Eq. (28) is then equivalently written as

$$c(\hat{x},\hat{t}) = \frac{M}{(\pi D \hat{t})^{1/2}} \exp\left[\frac{1}{2D}\left(V\hat{x} - \frac{\hat{x}^2}{2\hat{t}} - \frac{V^2 \hat{t}}{2}\right)\right] \qquad (29)$$

When $\beta=1$ and $\alpha=1$, the Hausdorff dimensions coincide with the Euclidean dimension so that the conventional dimensions of $D$ and $V$ in Eq. (28) become $[L^2/T]$ and $[L/T]$, respectively.

The advantage of Eq. (28) over Eq. (29) for parameter estimation is that data from the field or laboratories which are defined in the Euclidean space can be used to estimate the Hausdorff fractal space-time parameters, $\beta$ and $\alpha$. However, the final data on $D$ and $V$ need to be interpreted using Eq. (29) due to the fact that the dimensions of $D$ and $V$ in the Hausdorff space and time are $[L^2/T]$ and $[L/T]$, respectively. In other words, we use Eq. (28) to estimate the Hausdorff fractal parameters in the Euclidean space while interpreting their values in the Hausdorff space and time.

## 4. Methods for parameter estimation

### 4.1 Tritium as the tracer

The data of the scaled breakthrough curves used in this study were from the experiment given in (Pang and Hunt, 2001). In the experiment, tritiated water was injected into a column which was 8 m long with a 30 cm internal diameter. The column was filled with pea-gravel of uniform grain size. About 5 L of the solute containing tritium at 10,000 dpm/10ml was injected for 9 minutes. Samples were collected at 2 m, 4 m, 6 m and 8 m downstream from the injection point. The velocities of the pore-water were 28~34 m/day. At the end of the injection, the solute transported only 0.2-0.3 m downstream, which was much smaller than the sample collected at different positions. Thus, the tritium could have been injected as an instantaneous source. For more details about the experiment for the data cited here, the reader is referred to (Pang and Hunt, 2001).

### 4.2 Data analysis and curve-fitting

It should be noted that the reported flow rates observed in the first 2 hours were not very stable, and when the location is $x = 2$ m, the travel time after injection was 1.2 hours. Thus, the data of the scaled breakthrough curves were not very reliable at smaller values of $x$ due to disturbances at the point of injection and, in the present study, the tritium data at the further three locations $x = 4$ m, 6 m, and 8 m are investigated only. The solution of the solute transport equation for an instantaneous source presented in Eq. (28) is fitted to the scaled breakthrough curves reported in (Pang and Hunt, 2001). In Eq. (28), the time Hausdorff derivative order $\alpha$, space Hausdorff derivative order $\beta$, the dispersion coefficient $D$, the flow velocity $V$ were determined for all the three locations by using a nonlinear least square regression algorithm in MATLAB.

## 5. Results

Fig. 1 shows the fitting results of the scaled breakthrough curves at $x$ = 4 m, 6 m, and 8 m by using the solution of the Hausdorff derivative model (HDM) in Eq. (28). The results of the scale-dependent dispersion model (SDM) (Pang and Hunt, 2001), and the time-scale dependent dispersion model (TSDM) (Su et al., 2005) are also plotted in the figure. It should be pointed out that the calibrated constant dispersion model (CDM) with a constant diffusion coefficient and constant flow velocity gives almost similar results with the SDM. Thus, the results of the CDM are not provided here and more details can be found in (Pang and Hunt, 2001). The following analysis is concentrated on the peaks, early arrivals and heavy tails of the curves.

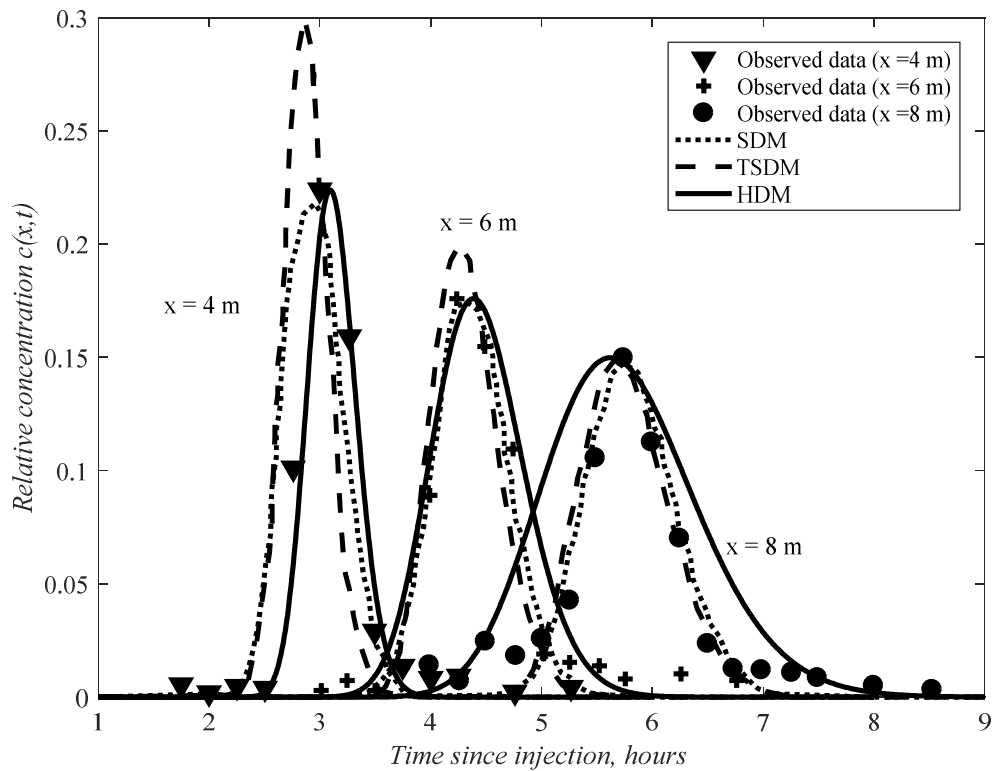

Fig. 1. Observed and simulated relative tritium concentrations at $x$ = 4 m, 6 m, 8 m using the solution of the Hausdorff derivative model (HDM) in Eq. (28), the scale-dependent dispersion model (SDM), and the time-scale dependent dispersion model (TSDM). The observed data are from (Pang and Hunt, 2001).

Compared to the observed relative concentration, the TSDM only provides good fits for the peak of the observed data toward the end of the column, *i.e.*, $x$ = 8 m, as given in Fig. 1, but overestimates the tritium concentration for smaller values of $x$. The HDM and the TSDM give equally good fits for the peaks of the scaled breakthrough curves in the three scenarios. For the central parts of the scaled breakthrough curves, the two models SDM and TSDM have better advantages over the HDM at larger values of $x$ for fitting the large values of relative concentration. More specifically, the HDM overestimates the relative concentrations for the central parts of the scaled breakthrough curves when the transport distance is large.

In Fig. 1, when the time after injection is small, the early arrival exists and is easier to be recognized compared with the case for $x = 4$ m. The early arrival of tritium dispersion can be successfully captured by the HDM for larger values of $x$. In the HDM, the value of the space Hausdorff derivative order $\beta$ determines the property of the early arrival. The smaller the values of $\beta$ the clearer early arrival occurs in the solute transport. For the recession part of the scaled breakthrough curves, it is observed from the three figures that the estimated results by the SDM and the TSDM decay faster than the observed data. The SDM and the TSDM underestimate the tritium concentrations for the tail. The HDM is capable of capturing the heavy tail. Similarly, the smaller values of $\alpha$ in the HDM, the longer the tail exists in the scaled breakthrough curves.

Table 1.  Estimated values of $\alpha$, $\beta$, $D$ and $V$ in Eq. (28)

| $x$, m | $\alpha$ | $\beta$ | $D$, $m^{2\beta}/h^{\alpha}$ | $V$, $m^{\beta}/h^{\alpha}$ |
|---|---|---|---|---|
| 4 | 0.74 | 0.80 | 0.0053 | 1.3125 |
| 6 | 0.58 | 0.65 | 0.0064 | 1.3575 |
| 8 | 0.47 | 0.55 | 0.0070 | 1.3917 |
| Average from 4 to 8 | 0.60 | 0.67 | 0.0062 | 1.3539 |

Table 1 summarizes the optimized values of $\alpha$, $\beta$, $D$ and $V$ for the above three locations. Considering the results presented in Table 1, the parameters $\alpha$ and $\beta$ characterize the transport processes along the travel path, which are consistent with the results illustrated in Fig. 1.

To examine the underlying relationships between the estimated parameters and the transport distances, Fig. 2 is prepared to show the variation of $\alpha$ and $\beta$ as the distance increases.

In Fig. 2, the time Hausdorff derivative order decreases linearly as the transport distance increases. The slope of the fitted straight line is -0.0675. It can be observed from Fig. 2 that the relationship between the space Hausdorff derivative order and the transport distance is also almost a straight line with the slope being -0.0625. Then the time and space Hausdorff derivative orders are also linearly correlated. By using the obtained relationships, for arbitrary transport distance from 4 m to 8 m, the values of the Hausdorff derivative orders at other locations can be determined by interpolation. However, the extrapolation of this relationship should be done with caution.

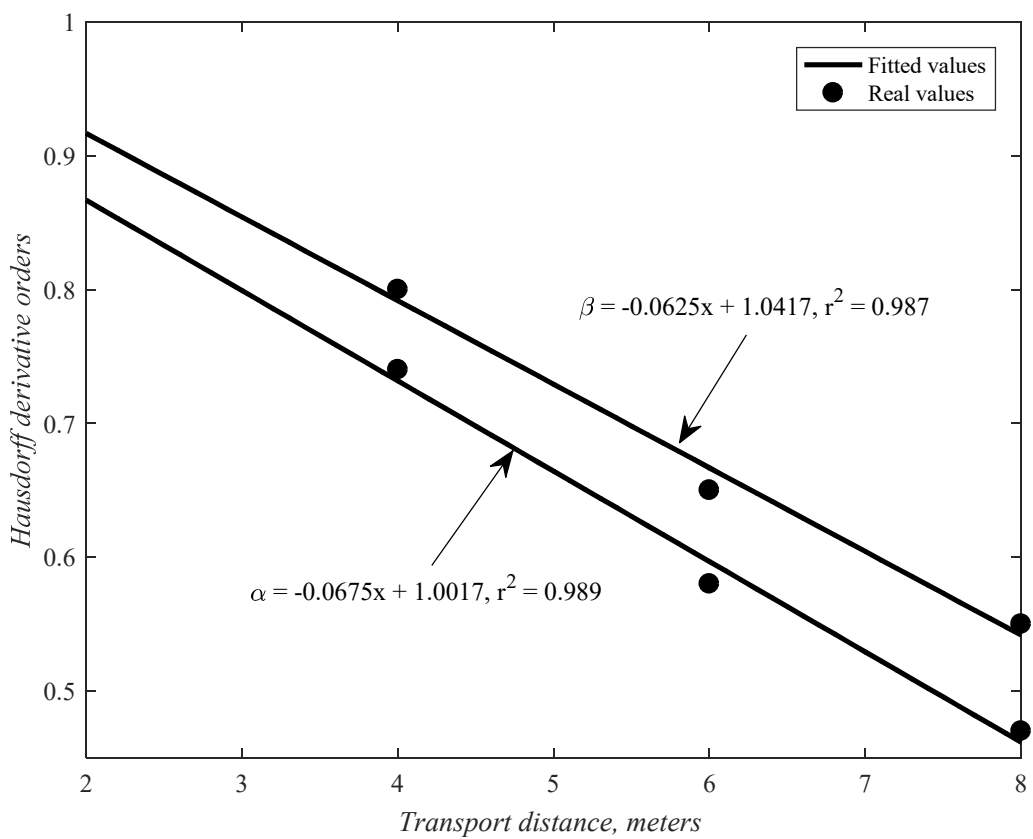

Fig. 2. The relationship between the time Hausdorff derivative order and the transport distance.

## 6. Discussion

The current study demonstrates that the equation of solute transport in heterogeneous porous media based on the Hausdorff fractal space can be satisfactorily used to describe solute transport in fractal porous media. The Hausdorff fractal model presented here can be transformed to the non-fractal advection-dispersion equation with a fractal time-space dependent dispersion coefficient. As an example of its applications, an explicit solution of the model with a constant diffusion coefficient and flow velocity is provided and simulates satisfactorily the scaled breakthrough curves for different solute travel distances.

The time and space Hausdorff derivatives are local and defined on the fractal metrics, which can be considered as a time-space power law scaling transform. The time and space Hausdorff derivative orders capture, respectively, the complexity in transport trajectory and transport scale. It is interesting to note that the estimated time and space Hausdorff derivative orders are linear functions of the transport distance. The Hausdorff derivative model provides a clear physical interpretation of the time- and space-dependent transport from the perspective of fractal metrics, which clearly defines the velocity and distance on fractals. Compared with the scale-dependent dispersion model and the time-scale dependent dispersion model, the Hausdorff fractal model better explains transport phenomena in porous media.

In this paper we derived one form of the solution of Eq. (9) with a constant diffusion coefficient and constant velocity for the case of an instantaneous source input. For solutions with different initial and boundary conditions, Nie *et al.* (2017) derived an explicit solution of an equation identical to Eq. (9) to model the transport of suspended sediment in unsteady flows where both the diffusion coefficient and flow velocity are constants. In reality the flow velocity and the dispersion coefficient of heterogeneous media usually vary, and they are often time- and scale-dependent such as in (Su et al., 2005) where Su *et al.* used the time- and space-dependent diffusion coefficient of the form $D(x,t) = D_0 x^m t^\lambda$. In future studies, solutions of Eq. (9) with a time- and scale-dependent diffusion coefficient would be investigated.

A simplified form of the model is one which has spatial fractal properties only such as Eq. (19) which is a result with $\alpha = 1$. Two potential strategies may be employed to explicitly derive the solutions of Eq. (19), which include the similarity solutions similar to those of the generalized Feller equation (GFE) (Lehnigk, 1993) and solutions based on integral transforms such as Laplace transform and/or Fourier- Laplace transforms.

It should be pointed out that another useful definition of fractal derivative also has clear geometrical mechanism (He, 2014). However, it is also a power law fractal metrics. For more complicated media, the fractal structure cannot be well described by the power law fractal metrics.

Instead, its structural metrics can be described by the generalized fractal metrics (GFM) in a non-Euclidean space as in (Chen, 2017),

$$\begin{cases} \Delta \hat{t} = \Delta T(t) \\ \Delta \hat{x} = \Delta Q(x) \end{cases} \quad (30)$$

where $T(t)$ and $Q(x)$ are structural functions. Eq. (30) becomes Eq. (2) when both $T(t)$ and $Q(x)$ are power functions.

In the context of structural metrics, the local structural derivative (Chen et al., 2016) is easily achievable,

$$\frac{ds}{dk(t)} = \lim_{t_1 \to t} \frac{s(t_1) - s(t)}{k(t_1 - t_0) - k(t - t_0)} \quad (31)$$

To capture the time and space complexity in a dispersive environment underlying the heterogeneous porous media, more general mathematical models are needed to represent the media complexity. To this extent, further investigations of GFM are required to demonstrate its applicability in solute transport in porous media.

## 7. Conclusions

This paper presents a new model based on the Hausdorff fractal concept. A simplified model with constant coefficients is demonstrated and verified using the tritium data in the form of observed breakthrough curves. The solution of the Hausdorff fractal model with the Hausdorff orders $\alpha$ and $\beta$ represents the fractal complexity of the media. The relationships between the parameters in the model and the travel distance are shown to vary, which is an indication that a physical interpretation of the model parameters is needed, and functions are needed for $\alpha$ and $\beta$.

Based on the foregoing results and discussion, the following conclusions are drawn:

1. The time and space Hausdorff fractal model is an alternative tool for characterizing solute transport in heterogeneous porous media. The new model better captures the early arrival and heavy tail of the breakthrough curves at different locations of the flow paths compared to the scale-dependent dispersion model and the time-scale dependent dispersion model with integer parameters.
2. The time and space Hausdorff derivative orders $\alpha$ and $\beta$ in the model incorporate the

heterogeneity of the porous media in diffusional dynamics, and can be respectively considered as measures of the early arrival and long tail in the breakthrough curves.

3. The time Hausdorff derivative order decreases linearly as the travel distance increases. The relationships between the Hausdorff derivative orders and the travel distance are almost a straight line, and the time and space Hausdorff derivative orders are linearly correlated.

4. The solution in Eq. (28) is used for estimating the Hausdorff fractal space-time parameters with the data defined in the Euclidean space. The estimated data on $D$ and $V$ need to be interpreted using Eq. (29) in the Hausdorff space and time so that the dimensions of $D$ and $V$ would be $\left[L^2/T\right]$ and $\left[L/T\right]$, respectively, which are consistent with the conventional units.

**Acknowledgments**

The work presented in this paper was supported by the National Natural Science Foundation of China (Nos. 11702085, 11772121), the Fundamental Research Funds for the Central Universities (No. 2017B01114), and the China Postdoctoral Science Foundation (No. 2018M630500).